%% file: casa.tex
\documentclass[twocolumn,linenumbers]{aastex63}
\usepackage{lineno}
\usepackage{hyperref}
\usepackage[utf8]{inputenc}
\usepackage{amsmath}
\usepackage{amsfonts}
\usepackage{enumerate}
\usepackage{ytableau}
\usepackage{graphicx}% Include figure files
\usepackage{indentfirst}
\usepackage[english]{babel}
\usepackage{overpic}
\usepackage[normal,footnotesize]{caption2}
\def\baselinestretch{1.25}
\usepackage{appendix}
\usepackage{amssymb}
\usepackage{amsthm}
\usepackage{epsfig,graphicx,epstopdf}

\newcommand{\gray}{$\gamma$-ray\ }
\newcommand{\grays}{$\gamma$-rays\ }

\newcommand{\U}[1]{\, \mathrm{#1}}
\newcommand{\uG}{\mu\rm G}
\newcommand{\Ep}{E_{\rm p}}

\submitjournal{ApJL}

\shorttitle{$\gamma$-ray spectrum of Cas A }
\shortauthors{LHAASO Collaboration}
\begin{document}

\title{Broadband $\gamma$-ray spectrum of supernova remnant Cassiopeia~A}

\nolinenumbers
\input{authors}

%\textcolor{red}{}

\begin{abstract}
\nolinenumbers
The core-collapse supernova remnant (SNR) Cassiopeia A (Cas~A) is  one of the brightest galactic radio sources with an angular radius of $\sim 2.5 \arcmin$. Although no extension of this source has been detected in the $\gamma$-ray band, using more than 1000 days of LHAASO data above $\sim 0.8$ TeV, we find that its spectrum is significantly softer than those obtained with Imaging Air Cherenkov Telescopes (IACTs) and its flux near $\sim 1$ TeV is about two times higher. In combination with analyses of more than 16 years of \textit{Fermi}-LAT data covering $0.1 \, \mathrm{GeV} - 1 \, \mathrm{TeV}$, we find that the spectrum above 30 GeV deviates significantly from a single power-law, and is best described by a smoothly broken power-law with a spectral index  of $1.90 \pm 0.15_\mathrm{stat}$ ($3.41 \pm 0.19_\mathrm{stat}$) below (above) a break energy of $0.63 \pm 0.21_\mathrm{stat} \, \mathrm{TeV}$. Given differences in the angular resolution of LHAASO-WCDA and IACTs, TeV $\gamma$-ray emission detected with LHAASO may have a significant contribution from regions surrounding the SNR illuminated by particles accelerated earlier, which, however, are treated as background by IACTs. Detailed modelling can be used to constrain acceleration processes of TeV particles in the early stage of SNR evolution.

\end{abstract}


\section{Introduction} 

Cassiopeia~A (Cas A, SNR G111.7-02.1) is one of the youngest supernova remnants (SNRs) in our Galaxy, originating from a core-collapse Type IIb supernova explosion approximately 350 years ago \citep{Reed1995, Krause2008, Fesen2006}. Located at a distance of 3.4 kpc, this source has a diameter of around 5 pc \citep{Kassim1995, Reed1995}. As one of the brightest galactic radio sources, Cas A has been extensively studied across various wavelengths from radio to X-rays, and up to ultra-high-energy $\gamma$-ray band \citep{2024ApJ...961L..43C}.

The bright radio emission, forming a circle with a radius of $\approx 1.7 \arcmin$, corresponds to the location where the ejecta interacts with the reverse shock \citep{Bell1975, Baars1977, Braun1987, Kassim1995}. In addition, a fainter radio emission extending up to a radius of $\approx 2.5 \arcmin$ has also been observed \citep{Delaney2014} and attributed to the forward shock. These radio emissions are produced via the synchrotron process by relativistic electrons. Synchrotron emission in the near-infrared (IR)  has also been detected at $2.2 \U{\mu m}$ (K-band) \citep{Gerardy2001, Rho2003, Jones2003},
and correlates well with the radio emission. Broadband spectral measurements from radio to IR exhibit significant curvature, suggesting very efficient particle acceleration and potential modifications in shock dynamics due to the back reaction of accelerated cosmic rays \citep{Rho2003, 2014ApJ...785..130Z}.

Non-thermal X-ray emission in the form of a narrow rim at the forward shock, detected with the \textit{Chandra} X-ray Observatory at $4-6 \U{keV}$, delineates the boundary of the remnant with a radius of $2.5 \arcmin \pm 0.2 \arcmin$ \citep{Gotthelf2001}. This emission is interpreted as synchrotron radiation from electrons accelerated to $\sim 40-60 \U{TeV}$ \citep{Gotthelf2001, Vink_MF}. Hard X-ray emission is also found in the reverse shock region, particularly in the western part \citep{Uchiyama2008, RSxray}. 
Analysis of ten years of \textit{INTEGRAL} data by \citet{Wang2016} revealed non-thermal X-ray continuum emission up to $220 \U{keV}$ that can be fitted with a smooth power-law without a high-energy cut-off. In addition to non-thermal X-ray emission, strong thermal X-ray component, primarily from line emission of shocked metal-rich ejecta \citep{Holt1994, Hwang2004}, is observed.

Cas~A was first detected in $\gamma$-rays at TeV energies by HEGRA stereoscopic Cherenkov telescope system \citep{2001A&A...370..112A}, showing a power-law photon spectrum with an index of $2.5 \pm 0.4_{stat} \pm 0.1_{sys}$ between $1 \U{TeV}$ and $10\U{TeV}$, and an integral flux above $1 \U{TeV}$ of $(5.8 \pm 1.2_{stat} \pm 1.2_{sys}) \times 10^{-13} \U{cm^{-2}\: s^{-1}}$, results later confirmed by the MAGIC \citep{Albert2007} and VERITAS \citep{Acciari2010} collaborations. In 2017, the MAGIC collaboration demonstrated that a power-law distribution with an exponential cut-off is favored over a single power-law distribution with a significance of 4.6 standard deviations with a spectral cut-off energy of $3.5_{-1.0}^{+1.6} \U{TeV}$ \citep{Ahnen2017}.
In the MeV-GeV range, Cas~A was initially reported by \textit{Fermi}-LAT based on one year data \citep{Abdo2010}. Subsequent detailed spectral analysis using 3.6 years of \textit{Fermi}-LAT observations revealed a significant spectral break at $1.72_{-0.89}^{+1.35} \U{GeV}$ \citep{Yuan2013, Saha_2014}, a feature confirmed with later analysis of $\sim 8$ years of \textit{Fermi}-LAT data by \citet{Ahnen2017}. 
Recently, a PL spectral index of $\Gamma = 2.17 \pm 0.02_{stat}$ and a cut-off energy of $2.3 \pm 0.5_{stat} \U{TeV}$ were inferred above this break with an updated analysis of VERITAS data covering 200 GeV - 10 TeV, and 10.8 years of \textit{Fermi}-LAT data covering 0.1-500 GeV \citep{Abeysekara2020}, suggesting that Cas~A could not be a PeVatron at its present age. Meanwhile, recent LHAASO-KM2A measurements \citep{2024ApJ...961L..43C} set stringent upper limits on the total energy of relativistic protons accelerated by Cas A. 

In this work, we describe LHAASO observations of Cas~A and update the GeV spectrum based on the 16.1 years \textit{Fermi}-LAT data. 
Multiwavelength data are then used to constrain models in the leptonic and/or hadronic scenarios for the $\gamma$-ray emission.

%%%%%%%%%%%%%%%%%%%%%%%%%%%%%%%%%%%%%%%%%%%%%%%%%%%%%%%%%%

\section{Observations}

%%%%%%%%%%%%%%%%%%%%%%%%%%%%%%%%%%%%%%%%%%%%%%%%%%%%%%%%%%
\subsection{LHAASO data analysis} \label{sec:data}

LHAASO, a multi-purpose and comprehensive extensive air shower (EAS) array, has been specifically designed to investigate cosmic rays (CRs) and $\gamma$-rays across a broad energy spectrum, ranging from sub-TeV to beyond 1 PeV. This state-of-the-art facility comprises three interconnected detector arrays: the kilometer-Square Array (KM2A) array, covering an area of 1.3 km$^2$, with unprecedented $\gamma$-ray detection sensitivity above 20 TeV; the Water Cherenkov Detector Array (WCDA), spanning 78,000 m$^2$, designed for TeV gamma-ray detection; and the Wide Field-of-view Cherenkov Telescopes Array (WFCTA), primarily focused on CR physics. The detailed information about the performance of these sub-arrays  can be seen in \citet{2019arXiv190502773C}.

For the data presented in this study, the full-array setup was employed, utilizing the WCDA data from March 5th, 2021, to July 31st, 2024, a total effective live time of approximately 1136.2 days. The event selection criteria are the same as in our previous work \citep{Aharonian_2021}. For the WCDA data, events are classified by the number of hits in seven analysis bins, i.e., 30-60, 60-100, 100-200, 200-300, 300-500, 500-800, $\geq$800, roughly corresponding to an energy range of 0.8 to 20 TeV. We also update the KM2A results for $10\ {\rm TeV} <E<1000$ TeV, using the full array data from July 20, 2021 to July 31st, 2024, a total effective observation time of 1064.9 days after data quality selection. The KM2A events are divided into bins with reconstructed energy width of 0.2 in log scale \citep{Aharonian_2021b}. In this paper, we present a survey encompassing a 4\arcdeg\ by 4\arcdeg\  region centered around  Cas~A. The sky map in celestial coordinates  is discretized in $0.1^\circ \times 0.1^\circ$ pixels.  We adopt the “direct integration method” \citep{Fleysher_2004} to estimate background in each pixels.

The test statistic (TS) is used to evaluate the significance of the source emission, defined by $TS = 2\ln({\mathcal L}_{s+b}/{\mathcal L}_b)$, where ${\mathcal L}_{s+b}$  and ${\mathcal L}_b$ represent the likelihoods for the signal plus background  hypothesis and the background only hypothesis, respectively \citep{2024ApJS..271...25C}. The spectral energy distribution (SED) of the gamma-ray emission is calculated using the forward-folding method, assuming that the energy spectrum follows a power law (PL). The energy spectrum parameters can be obtained by maximizing the likelihood value \citep{2024ApJS..271...25C}.

\begin{figure}
    \centering
    \includegraphics[width=0.48\linewidth]{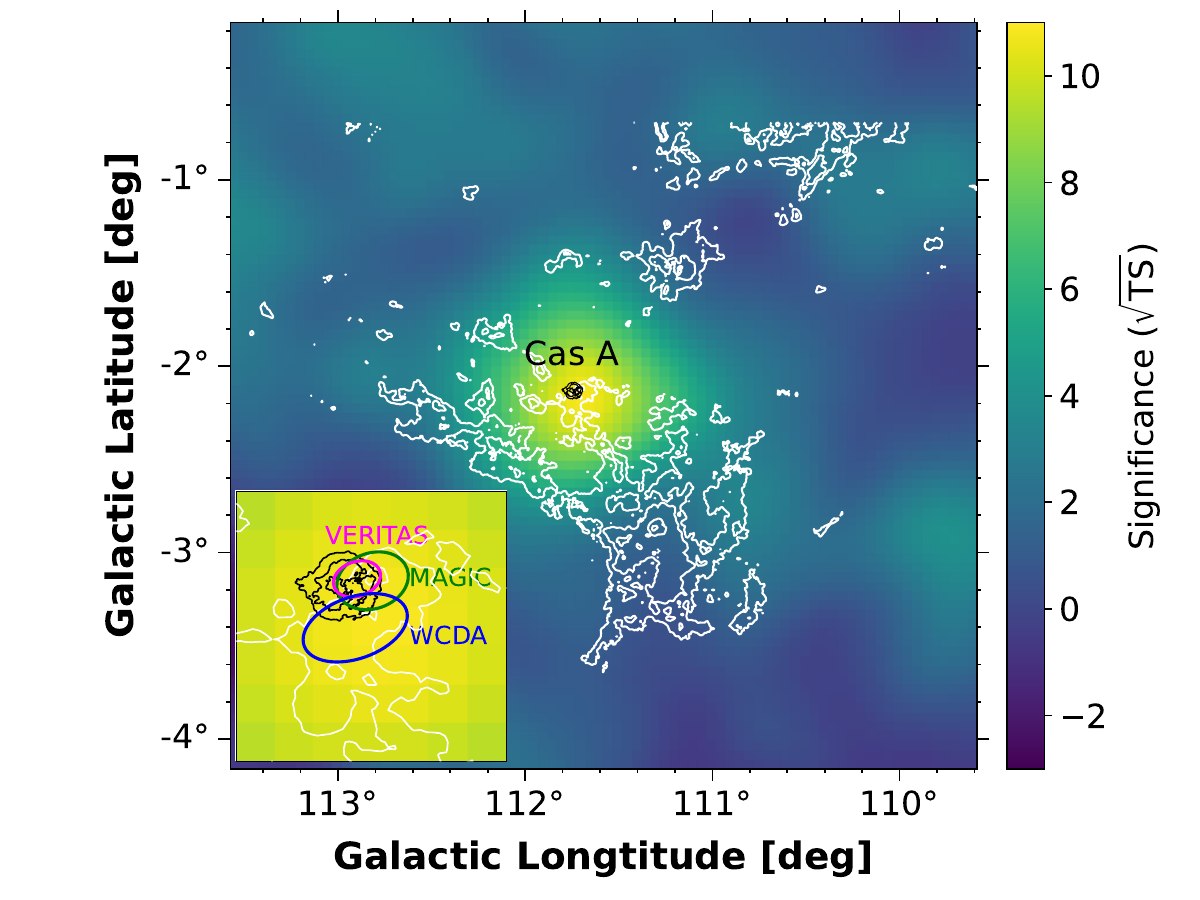}
    \includegraphics[width=0.49\linewidth]{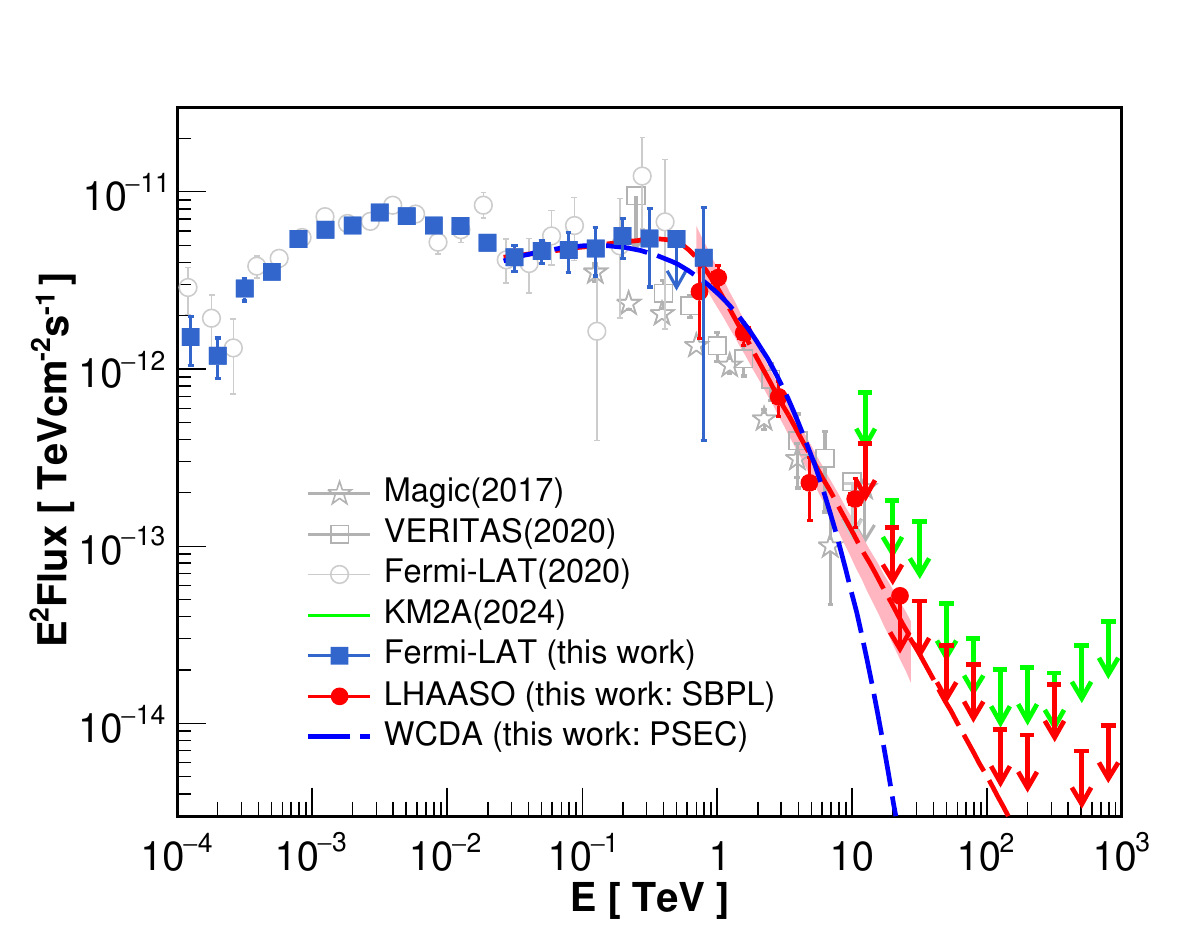}
    \caption{Left: The LHAASO TS map in the energy range of 0.8~TeV to 25~TeV in the $4^{\circ}\times4^{\circ} $ region around Cas A, the colorbar shows the $\sqrt{\rm TS}$ value. The white contours are derived from $^{12}\mathrm{CO}$ line observations \citep{ma19}. The black X-ray contours mark the position of Cas A. In the insert at the left-bottom corner, the positions with $1\sigma$ uncertainty of LHAASO, VERITAS, and MAGIC are also shown in blue, magenta, and green ellipses, respectively. Right: The $\gamma$-ray spectra of Cas A.   The data points are from Fermi-LAT (solid blue square) and  LHAASO  in this work (SBPL: Red butterfly, solid circles  and upper limits; PSEC: Blue dashed line). The MAGIC flux points in gray open pentagram \citep{Ahnen2017}, the VERITAS/Fermi-LAT  flux points in gray filled square/open circle \citep{Abeysekara2020}, and the green upper limits of KM2A \citep{2024ApJ...961L..43C} are also shown.}
    \label{fig:sed_gamma}
\end{figure}

In the energy range of 0.8~TeV to 25~TeV, the {LHAASO} significance  map of Cas~A region is shown in the left panel of Figure \ref{fig:sed_gamma}. The $\gamma$-ray spatial distribution is well described by a point-like source with a TS of 120.7 (corresponding to $\sim 11.0\,\sigma$). The $\Delta \rm{TS}$ value between the point-like model and 2-D Gaussian model is smaller than 1.4. The best-fit position of the source is (RA = $350.87^\circ \pm 0.07^\circ_{stat}$, Dec = $58.75^\circ \pm 0.04^\circ_{stat}$) assuming a power law (PL) spectrum:  ${dN(E)}/{dE}=N_0({E}/{\rm 3TeV})^{-\varGamma}$ (Table \ref{tab:Values}).  
For the WCDA data, an investigation of the pointing systematic error has been conducted using three point sources: Crab Nebula, Mrk 421, and Mrk 501. The largest error is from Mrk 421 near $0.039^\circ$, which is taken as the pointing error of the LHAASO-WCDA array in this work. The positions (listed in Table~\ref{tab:Values}) determined with the LHAASO are slightly shifted southward (see insert of the left panel of Fig.\ref{fig:sed_gamma}) and are consistent with \textit{Fermi}-LAT \citet{Yuan2013, abdollahi2020fermi} and VERITAS \citep{Abeysekara2020} within statistical and systematic uncertainties, as well as with the center of the remnant.

%%%%%%%%%%%%%%%%%%%%%%%%%%%%%%%%%%%%%%%%%%%%%%%%%%%%%%%%%%
%%%%%%%%%%%%%%%%%%%%%%%%%%%%%%%%%%%%%%%%%%%%%%%%%%%%%%%%%%
\subsection{\textit{Fermi}-LAT data analysis} 

We analyzed the Fermi-LAT data using the EasyFermi software \citep{2022A&C....4000609D,WoodFermipy2017,2023A&A...678A.157D,2018AJ....156..123A,2013PASP..125..306F}. The dataset covers a period of 16.1 years of \textit{Pass8R3} \textit{Fermi-LAT} data, spanning from August 4, 2008, to September 10, 2024. We selected events from a $15^{\circ} \times 15^{\circ}$ region centered on the position of Cas~A in the energy range from $100\U{MeV}$ to $1000\U{GeV}$. We binned the data with a pixel size of $0.1\degr$ and five energy bins per decade. To reduce the $\gamma$-ray contamination from the Earth limb, a maximum zenith angle of 90$\degr$ is set. The event class $P8R3\_SOURCE$ (“evclass=128”), and corresponding instrument response functions (IRFs) P8R3\_SOURCE\_V3 is used in our analysis. For background model, we include the diffuse Galactic interstellar emission (IEM, $gll\_iem\_v07.fits$),  isotropic emission (``$iso\_P8R3\_SOURCE\_V3\_v1.txt$'' ) and all sources listed in the fourth \textit{Fermi}-LAT catalog \citep{abdollahi2020fermi}. 

For the 16.1-year \textit{Fermi}-LAT observation of Cas A, the source is detected with TS = 7855 in the whole energy band (corresponding to a significance of $\sim88.6\sigma$) by using a log-parabola (LP) function: 
\begin{equation}
    \label{eq:lp}
    {dN(E)}/{dE}=N_0({E}/{\rm 2.23GeV})^{-\varGamma-\beta {\rm ln} (E/{\rm 2.23GeV})},
\end{equation}
where $\varGamma$ is the photon index of the spectrum and $\beta$ is the coefficient reflecting the spectral curvature.  In this work, the best-fit source position of (RA $= 350.8611^\circ \pm 0.0022^\circ_{stat}$, Dec $= 58.8227^\circ \pm 0.0023^\circ_{stat}$) is consistent with that in the 4FGL-DR4 \citep{abdollahi2020fermi} with RA $= 350.8613^\circ$ and Dec $= 58.8173 ^\circ$, and the global fitting gives a photon index of $\varGamma=1.808 \pm 0.018_\mathrm{stat}$ and $\beta=0.097 \pm 0.009_\mathrm{stat}$. 
To calculate the SED, the whole energy range was divided into 20 logarithmically even-spaced bins and the normalization of all sources within 5 degree are freed.
The derived SED is shown by the blue filled squares in the right panel of Figure \ref{fig:sed_gamma}.

\subsection{Joint results for LHAASO and Fermi-LAT} 

We combined the LHAASO and \textit{Fermi}-LAT observed spectral energy distribution results, as shown in the right panel of Figure \ref{fig:sed_gamma}. In the energy range from 0.8 to 20 TeV, the flux level measured with LHAASO is consistent with \textit{Fermi}-LAT flux points above 30 GeV. Above 30 GeV, the joint spectrum from LHAASO and \textit{Fermi}-LAT deviates significantly from a simple PL, and is best described by a power-law with a sub-exponential cutoff (PSEC): 
\begin{equation}
	\label{eq:pl}
	{dN(E)}/{dE}=N_0({E}/{\rm 3TeV})^{-\varGamma}{\rm exp}(-(E/{E_{\rm cut})}^{0.5}),
\end{equation}
where $E_{\rm cut}$ is the cutoff energy, $\varGamma$ is the photon index of the spectrum, or a smoothly broken power-law (SBPL):
%SBPL function is defined as 
\begin{equation}
	\label{eq:sbpl}
	{dN(E)}/{dE}=N_0 \left(E/3\,{\rm TeV}\right)^{-\varGamma_1}\left[1+(E/E_{\rm br})^{s}\right]^{(\varGamma_1-\varGamma_2)/s},
\end{equation}
where $E_{\rm br}$ is the break energy, $\varGamma_1$ and $\varGamma_2$ are the photon indices above and below the break, respectively, and $s$ characterizes the smoothness of the break (fixed at 5 in this paper). 

To minimize the influence of the spectral model on the low-energy range of LHAASO and to reduce the number of model parameters for the joint analysis, 1) we fitted the combined Fermi data points above 30 GeV and the LHAASO-WCDA data points (from PL model) to obtain the initial parameters. For the SBPL, the results are $\varGamma_1 = 1.96 \pm 0.12 \, \mathrm{stat}$, $\varGamma_2 = 3.33 \pm 0.17 \, \mathrm{stat}$, and $E_{\rm br} = 0.62 \pm 0.18 \, \mathrm{stat} \, \mathrm{TeV}$. Conversely, for the PSEC, the values obtained are $\varGamma = 1.61 \pm 0.16 \, \mathrm{stat}$ and $E_{\rm br} = 0.18 \pm 0.08 \, \mathrm{stat} \, \mathrm{TeV}$.
2) The LHAASO observations, including both WCDA and KM2A, were then re-fitted with fixed $\varGamma_1 = 1.96 $ and $E_{\rm br}=0.62$~TeV for the SBPL, and $\varGamma=1.61$ and $E_{\rm cut}=0.18$~TeV for the PSEC to obtain the data points.
3) We re-fitted the Fermi data points above 30 GeV along with the LHAASO-WCDA data points to derive the final joint fitting parameters  (Table~\ref{tab:Values}). The new joint fitting parameters are consistent with the ones obtained from step 1 within statistical uncertainties so we stop the iterative process.  Finally, we updated the LHAASO data points by fixing the parameters to the ones from the final joint fit,which are shown in the right panel of Figure \ref{fig:sed_gamma}. The red upper limits for the KM2A are obtained with the SBPL spectral model in this work, while the spectral model was fixed to a power-law function with an index of 2.8 in the previous KM2A analysis \citep{2024ApJ...961L..43C}.  

Near 1 TeV, the fluxes measured with LHAASO is about two times those of the MAGIC and the VERITAS. The variation of the atmosphere condition during the operation affects the detection efficiency, which is not fully considered in the simulation. For point-like sources, the systematic uncertainties of the LHAASO-WCDA flux can be as large as 8.0\% \citep{2024ApJS..271...25C}, it also includes effects of long-term water quality changes.  In this analysis, we also conducted a comparison between the PSEC and SBPL models over the energy range of 0.8 TeV to 8 TeV. Specifically, we selected 10 evenly spaced points in $\log_{10}(E)$. We calculated the absolute differences for each point and then averaged these differences to determine the systematic error between these two models. The resulting systematic error is 15.9\%. Utilizing the same approach as described above, different spatial and spectral models of the diffuse  Galactic $\gamma$-ray emission give rise to a flux uncertainty of 11.0\%. Since Cas A is not resolved by all these observations, the flux differences are likely caused by differences in the spatial resolution of these observations if Cas A is surrounded by a dim $\gamma$-ray halo. IACTs have a better resolution than the LHAASO-WCDA and treated the surrounding regions as background. Therefore some emission from the halo may be treated as background by IACTs, giving rise to a lower flux. Observations with better sensitivity will be able to address this issue.

A PL function fit to the spectrum above 30 GeV yielded a $\chi^2$ of 54.9, while the PSEC or SBPL function yielded a $\chi^2$  of 10.5 or 5.0, respectively. According to the $AIC_{c}$\citep{2007MNRAS.377L..74L},
\begin{equation}
	\label{eq3}
	AIC_c=\chi^2 + 2k + \frac{2k(k+1)}{n- k -1},
\end{equation}
where $n$ is the number of data points and $k$ is the number of parameters of the corresponding model, the model with the smallest $AIC_{c}$ provides the best fit to the data. 
The SBPL ($AIC_{c}$=18.0) and the PSEC ($AIC_{c}$=19.5) are better than the PL ($AIC_{c}$=60.1), implying %that the index have change 
a spectral curvature near 1 TeV. $\Delta$ $AIC_{c}$ between SBPL and PSEC is 1.5, suggesting that they are statistically comparable.

%%%%%%%%%%%%%%%%%%%%%%%%%%%%%%%%%%%%%%%%%%%%%%%%%%%%%%%%%%
\section{Discussion}

The broad-band gamma-ray spectrum shown in Figure \ref{fig:sed_gamma} indicates a significant bump structure above 30 GeV.
To explore the nature of this $\gamma$-ray feature, we first fit the whole $\gamma$-ray spectrum ($\ge$100 MeV) with a simple hadronic one-zone emission model following previous studies \citep[e.g.,][]{casa_magic,Abeysekara2020}.
%first ignore the bump-like feature around TeV and 
% using the hadronic cross sections presented in \citep{Kafexhiu14} 
Considering the break around GeV detected by {\sl Fermi}-LAT \citep{Yuan2013,Abeysekara2020}, we assume that the particle distribution in the momentum space has a broken power-law form with a high-energy cutoff,  which is converted to the total particle energy ($E_{\rm p}$) space as
\begin{align}
\nonumber
 \frac{d N}{d\Ep} & \propto \frac{dN}{dp} \frac{dp}{d\Ep} \propto 4\pi p^2 f(p) \frac{dp}{d\Ep} \\ 
& \propto p^{-\alpha_1} 
 % \times \prod_{i=1}^n 
 \left[ 1+\left( \frac{p}{p_{\rm b}} \right)^2 \right]^{\frac{\alpha_1-\alpha_{2}}{2}} 
 % \times
 \mathrm{exp}\left(-\frac{p}{p_\mathrm{c} } \right)
 \frac{\Ep}{p}
 \label{eq:dis}
\end{align}
where $p$ is the momentum of protons, $p_{\rm b}$ is the break momentum, $p_{\rm c}$ corresponds to the high-energy cutoff, $\alpha_1$ and $\alpha_{2}$ are the power-law indexes below and above the break, respectively. 
The density of target gas is taken as $3.6\ \mathrm{cm}^{-3}$, which is obtained from the ambient gas density upstream of the forward shock accounting for the compression ratio of four expected for strong shocks \citep[$n_0=0.9\ \mathrm{cm}^{-3}$,][]{Lee2014}.
%For $\alpha_1=\alpha_{2}$, there is no break, corresponding to the power-law Model (M1). Otherwise, it is the broken power-law Model (M1).

\begin{figure}
    \centering
    \includegraphics[width=0.49\linewidth]{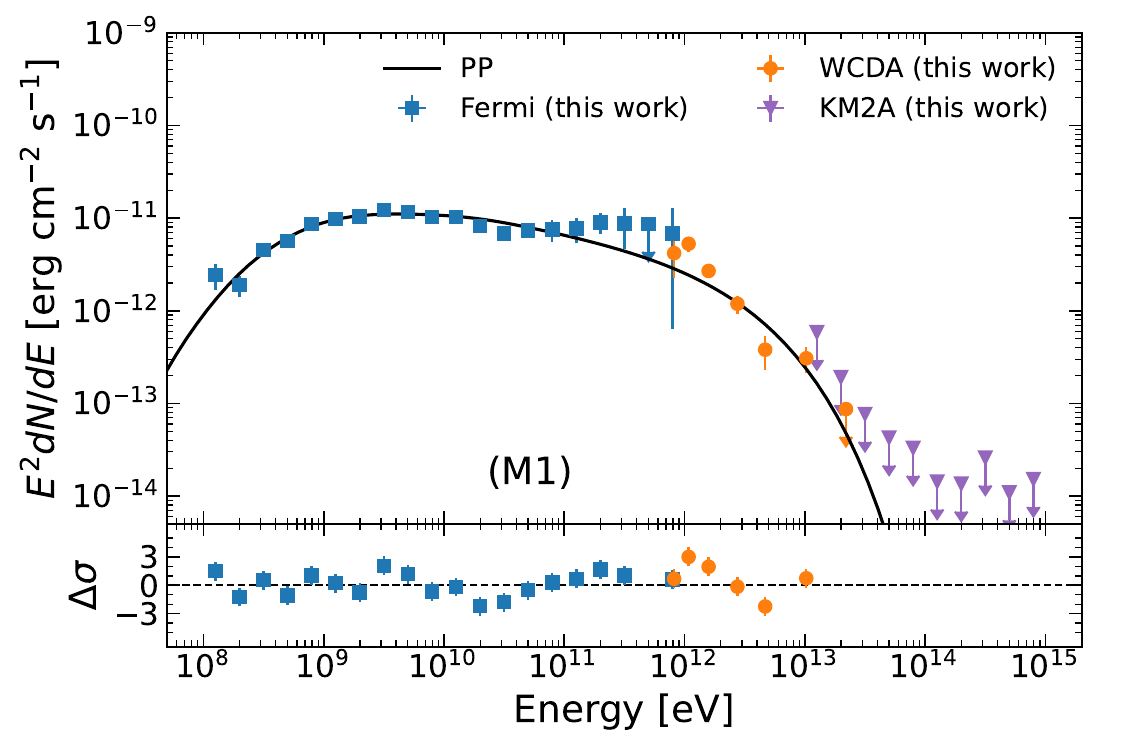}
    \includegraphics[width=0.49\linewidth]{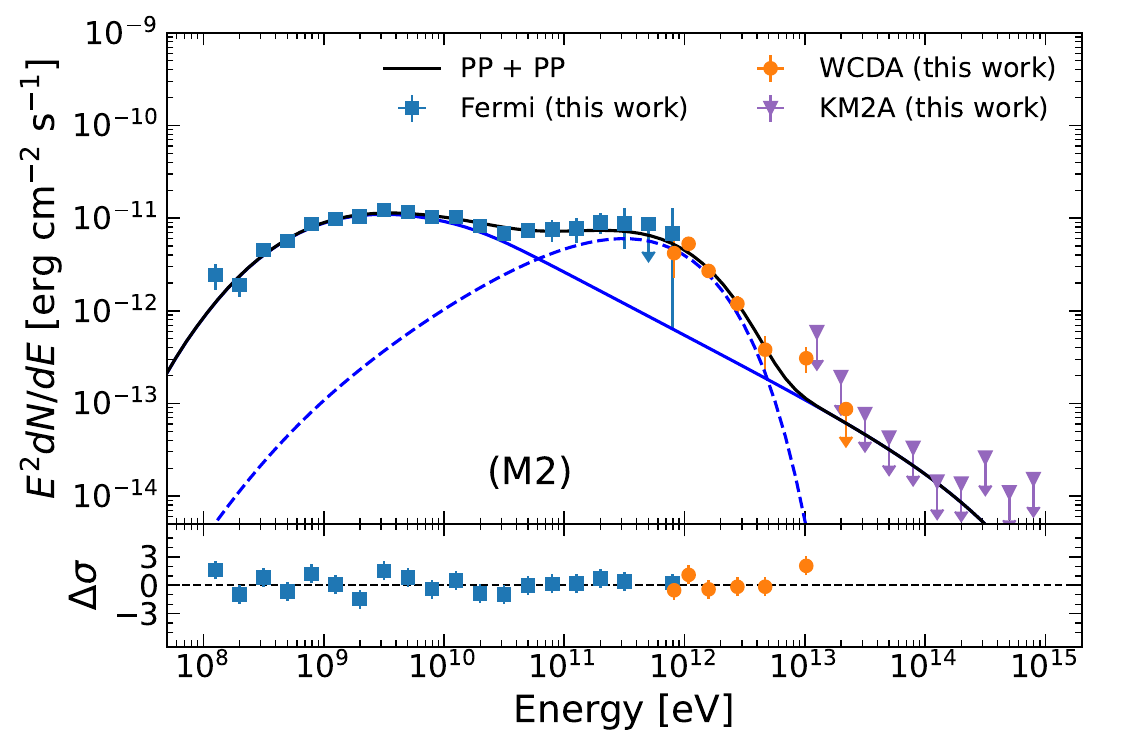}
    \includegraphics[width=0.49\linewidth]{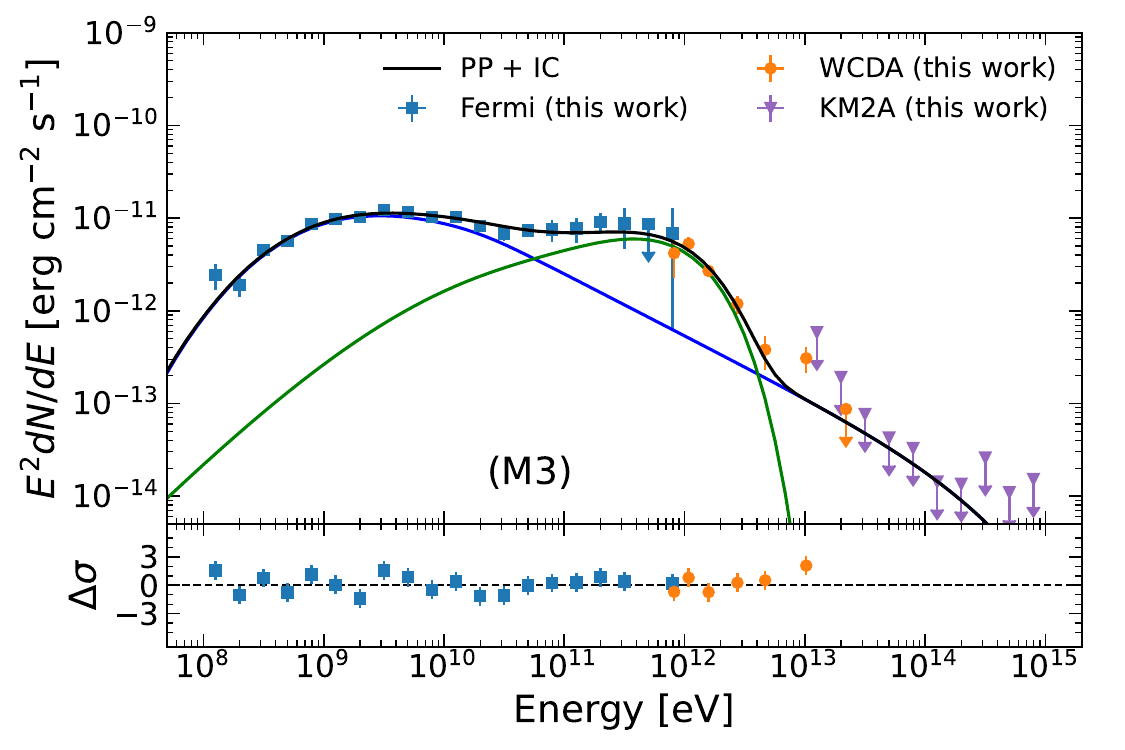}
    \includegraphics[width=0.49\linewidth]{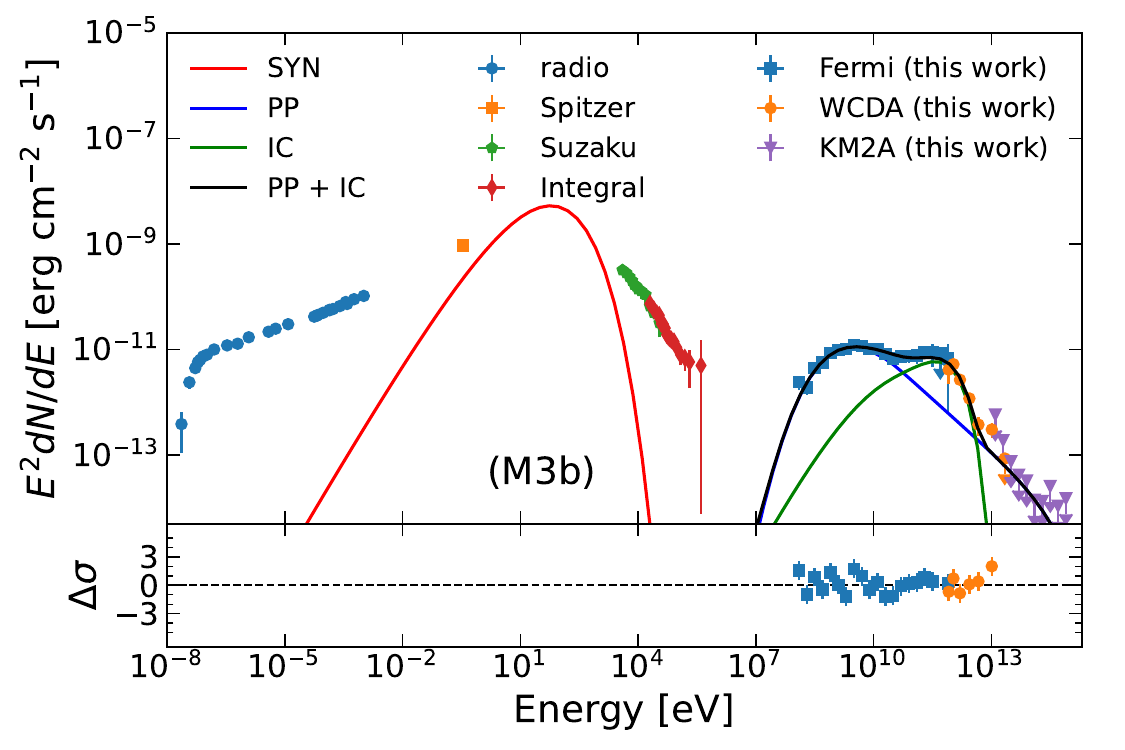}
    \caption{Best-fit {\gray} SEDs of SNR Cas~A for Models 
    M1 (Top Left), M2(Top Right), M3 (Bottom Left). Bottom Right: the same as M3 but shows the synchrotron spectrum with a magnetic field of $200\,\uG$.
    Also shown are the radio data (light blue filled circle) given in \citet{Vinyaikin2014}, infrared data from IRAC 3.6 $\mu \mathrm{m}$ \citep[orange filled square;][]{DeLooze2017}, X-ray data from Suzaku \citep[green filled pentagon;][]{Maeda2009} and INTEGRAL-IBIS \citep[red filled diamond;][]{Wang2016}.
    }
    \label{fig:sed1}
\end{figure}

Given above particle distribution and density (Model M1), we employ the PYTHON package Naima \citep{naima} with the hadronic cross sections presented in \citep{Kafexhiu14} to fit the $\gamma$-ray spectra and constrain the model parameter.
If all parameters are free, the posterior distribution of the index below the break approximately has the one-side Gaussian shape with $\alpha_1 < 1.28$ at 95\% percentile for the prior $\alpha_1 \in $~[0, 2.0]. But for the following models M2-M3b, the case will become worse due to lack of constraint on the index above the break. 
In fact, the index can not be well constrained if it is harder than the index of the $\gamma$-ray spectrum produced by the mono-energy protons of $\sim$0.6--0.7.
Therefore, we fix the index below the break as the median value $\alpha_1 = 1.0$ of the range from 0.7 to 1.3 for all models, which has limited impact on the results.
The fitted SED are displayed in Figure~\ref{fig:sed1} and the corresponding parameters including Bayesian information criterion \citep[BIC,][]{Kass1995,2007MNRAS.377L..74L} are listed in Table~\ref{tab:had_lep_param}.

Although M1 can fit the GeV data, there are some excesses near 1 TeV. Thus we add a second $\gamma$-ray component that can have a hadronic (M2) or leptonic (M3) origin for the TeV $\gamma$-ray emission.
The parent particle distribution of the second component is assumed to be a power law with an exponential cutoff $dN/dp_{\rm e/p}\propto p_{\rm e/p}^{\alpha_{\rm e/p}}\mathrm{exp}(-p/p_{\rm e/p})$.
For the leptonic process, we also use Naima to calculate the synchrotron and inverse Compton (IC) spectra with the formula presented in \citet{naima.syn} and \citet{naima.ics}, respectively.
In the IC process, the seed photon fields include the cosmic microwave background (CMB) and a far infrared field with a temperature of 100 K and an energy density of 2 ${\rm eV\ cm^{-3}}$ \citep{Mezger1986,Atoyan2000.TA}.
In models M2 and M3, the index above the break and the cutoff for the first component can not be simultaneously well constrained.
In the fitting process, the cutoff momentum is fixed at $p_{\rm c}=3\ {\rm PeV\ c^{-1}}$ \footnote{Any value of $p_{\rm c}$ above $\sim 30\ {\rm TeV\ c^{-1}} $ leads to essential the same fitting results.} although there is still a lack of conclusive evidence of SNRs as PeV accelerators, where $c$ is the speed of light, and the index $\alpha_2$ is set with prior in the range [2.0, 4.0]. 
According to the BIC, models M2 and M3 are much better than M1, suggesting that the second component is indeed necessary to explain the spectral bump around 1 TeV.

$\Delta$BIC between M2 and M3 is less than 2, suggesting they are statistically comparable.
However the spectral indexes of the emitting particle distribution 
$\alpha<1.6$, which is difficult to reproduce with the diffusive shock acceleration theory \citep{Malkov2001, 2014ApJ...785..130Z}.
On the other hand, considering the transport processes, the hadronic process may produce a hard $\gamma$-ray spectrum if the remnant is evolving in a clumpy medium \citep{Inoue2012, Gabici2014}.
This mechanism may also explain the differences between LHAASO observations and measurements with IACTs if particles escaping from the SNR and illuminating the surrounding background gas can have a significant contribution to the TeV emission.
The emission surrounding the SNR is treated as the background with IACTs. The WCDA observations on the other hand cannot separate it from those produced within the SNR.
Nevertheless based on the CO analysis, it was suggested that Cas A is not associated with molecular clouds \citep{Zhou18}. 
%Thus, the leptonic process are favoured by the current observations, although the hadronic origin can not be fully ruled out.

If the bump originates from the leptonic process, the same electron population should produce emission from radio to X-ray via the synchrotron process (M3b) assuming an
% This synchrotron emission with the
average magnetic field $B=200\, \uG$. 
As can be seen, this lepto-hadronic model (M3b) can not well fit the broadband data of Cas A, implying a second synchrotron component to explain the radio and hard X-ray data.

\begin{figure}
    \centering
    \includegraphics[width=0.69\linewidth]{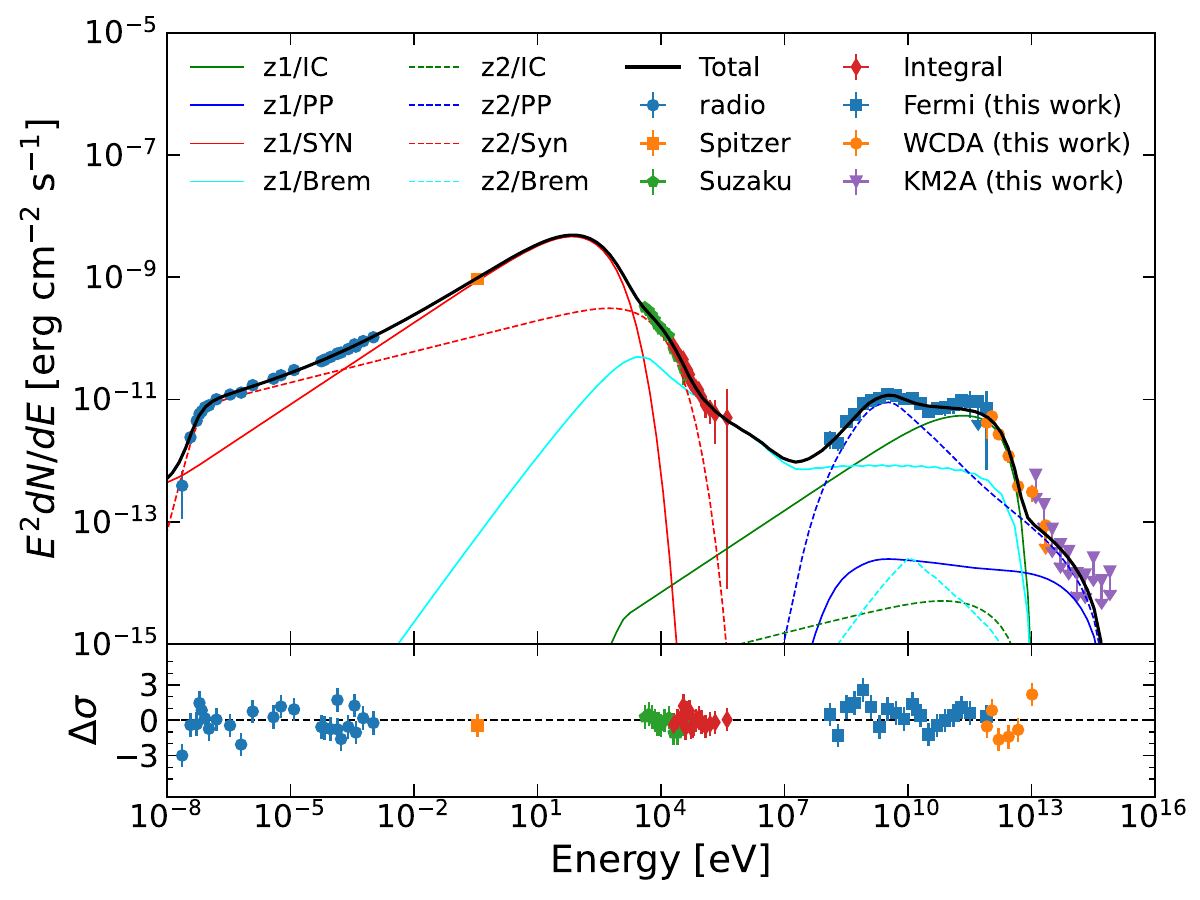}
    \caption{Broadband SEDs of SNR Cas~A fitted with 
    a two-zone model. The black solid line represents the total emission from zones 1 (solid) and 2 (dashed) with various components: synchrotron (red), inverse Compton (green), bremsstrahlung (cyan) and p-p collision (blue). Also shown are the radio data (light blue filled circle) given in \citet{Vinyaikin2014}, infrared data from IRAC 3.6 $\mu \mathrm{m}$ \citep[orange filled square;][]{DeLooze2017}, X-ray data from Suzaku \citep[green filled pentagon;][]{Maeda2009} and INTEGRAL-IBIS \citep[red filled diamond;][]{Wang2016}. }
    \label{fig:sed2}
\end{figure}

If the emission zone is not uniform, one may consider
multiple component emission models.
Based on observational features of Cas A, some two-zone models  have been proposed to explain the broadband spectra \citep{Atoyan2000.TA,Zhang2019.casa,liu22,Zhan2022.casa}.
In the model of \citet{Zhang2019.casa}, the emission zone is divided into two regions based on the model proposed by \citet{Atoyan2000.TA} and features in X-ray band: zone 1 consists of the outer ``thin rim" (the forward shocks) and the diffuse region immediately behind the forward shock. Zone 2 includes the rest of emission regions. 
\citet{liu22} improved this model by including the bremsstrahlung process of non-relativistic electrons.
%which contributes the hard X-rays.
Based on the asymmetric profile, \citet{Zhan2022.casa} proposed a two-zone model containing the jet-like structure (zone 2) superposed onto an expanding spherical shock (zone 1).
The synchrotron emission from jet-like structure dominated the $\sim$100 keV X-rays.
As an example, the case B of the two-zone model presented in \citet{liu22} is used to fit the broadband data as shown in Figure~\ref{fig:sed2}.
The fitted parameters are almost the same as that in \citet{liu22} except for a factor of 2 increase in the total energy of electrons in zone 1 with a hard distribution to fit the enhanced fluxes near 1 TeV. The magnetic field in this zone is also reduced slightly to produce a nearly identical synchrotron spectrum. As can be seen from the residuals in the lower panel, the model spectrum is slightly softer than the observed one below $\sim 1$ TeV, which is consistent with M3 with a harder electron distribution.

It is evident that simple emission models not considering transport effects may not explain the spectral bump above 30 GeV due to the hard distribution of emitting particles in M2 and M3. It is difficult to produce a spectral index smaller than 2 with  conventional acceleration models. In the diffusive shock acceleration theory, the overall distribution of accelerated particle can be hard in the upstream if the diffusion coefficient increases with energy. Therefore one may attribute these hard spectra to particles in the upstream of shocks. Escaping particles also have a hard spectrum in the standard particle acceleration model in SNR \citep{2014ApJ...785..130Z}. Interestingly, such kind emission has been seen in TeV bright SNR RX J1713.7-3946 \citep{2009A&A...505..157A}.
In the hadronic scenario, the inferred total energy of the emitting ions is about 4 times higher than that given in Table \ref{tab:had_lep_param} for the low density in the upstream. The corresponding acceleration efficiency of converting the shock kinetic energy into relativistic particles can be higher than 10\%. The spectral bump therefore may be attributed to spectral hardening below the spectral cutoff in the context of nonlinear diffusive shock acceleration \citep{2014ApJ...785..130Z, Malkov2001}. Detailed modelling is warranted \citep{2024ApJ...961L..43C}. 

\section{Conclusion}

By analyzing Fermi-LAT and LHAASO observations of Cas A, we have found a new spectral feature above 30 GeV. The spectrum of this feature is very soft above 1 TeV. Although both a sub-exponential cutoff and a soft power-law model can fit the LHAASO data, they predict drastically different spectra above 10 TeV and imply distinct sub-PeV particle acceleration processes \citep{2024ApJ...961L..43C}. Future spectral measurements may distinguish these two models. Although a two-zone emission model gives a reasonable fit to the broadband SED, the model spectrum is softer than the observed one between 30 GeV and 1 TeV. The hard emitting particle distribution inferred from detailed fitting of the $\gamma$-ray spectrum may be attributed to emission in the shock upstream, ion transport in an extended clumpy medium, and/or nonlinear diffusive shock particle acceleration processes. More detailed modelling is needed. Future observations with the IACT and/or LHAASO with a better angular resolution and sensitivity above 10 TeV {\grays} will be very helpful to reveal the nature of the spectral bump.

\acknowledgments
We would like to thank all staff members who work at the LHAASO site above 4400 meter above 
the sea level year-round to maintain the detector and keep the water recycling system, 
electricity power supply and other components of the experiment operating smoothly. We are 
grateful to Chengdu Management Committee of Tianfu New Area for the constant financial support 
for research with LHAASO data. We appreciate the computing and data service support provided 
by the National High Energy Physics Data Center for the data analysis in this paper. 
This work is supported by the following grants: the National Key R$\&$D program of China under the grant 2024YFA1611403, the Department of Science and Technology of Sichuan Province, China (Nos. 2024ZYD0111, 24NSFSC2319, 2024NSFJQ0060), the National Natural Science Foundation of China (Nos. 12220101003, 12173039, 12393851, 12393852, 12393853, 12393854, 12205314, 12105301, 12305120, 12261160362, 12105294, U1931201, 12375107, 12342502),  the Project for Young Scientists in Basic Research of Chinese Academy of Sciences (No. YSBR-061),  the National Science and Technology Development Agency (NSTDA) of Thailand, and the National Research Council of Thailand (NRCT) 
under the High-Potential Research Team Grant Program (N42A650868).

\bibliography{casa}{}
\bibliographystyle{aasjournal}

\begin{table}[th]
\centering
\caption{Best-fit values for the LHAASO data points and Fermi data points above 30 GeV. }
\label{tab:Values}
\begin{tabular}{cccc}
\hline
\hline
Spectral Types & PL$^a$ & PSEC & SBPL \\
RA & $350.87^\circ \pm 0.07^\circ_{stat}$ & $350.87^\circ$ & $350.87^\circ $\\
Dec & $58.75^\circ \pm 0.04^\circ_{stat}$ & $58.75^\circ $ & $58.75^\circ $\\
Flux@3 TeV/$\mathrm{TeV^{-1}cm^{-2}s^{-1}}$
%$N_0$/$\mathrm{TeV^{-1}cm^{-2}s^{-1}}
& $(6.64 \pm 0.84\, \mathrm{stat})\times 10^{-14}$ & $(9.01 \pm 0.85\, \mathrm{stat})\times 10^{-14}$ & $(6.99 \pm 1.35\, \mathrm{stat})\times 10^{-14}$ \\
$\varGamma$ or $\varGamma_1$/$\varGamma_2$ & $3.30 \pm 0.08\, \mathrm{stat}$ & $1.60 \pm 0.17\, \mathrm{stat}$ & $1.90\pm 0.15\, \mathrm{stat} / 3.41\pm 0.19\, \mathrm{stat} $ \\
$E_{\rm cut}$ or $E_{\rm br}$   &  & $0.20 \pm 0.09\, \mathrm{stat}$ TeV & $0.63 \pm 0.21\, \mathrm{stat}$ TeV  \\
\hline
\end{tabular}
\tablenotetext{a}{Best-fit values obtained exclusively for LHAASO data using the PL model. }
\end{table}

\begin{deluxetable}{ccccccccc}[htb]
\tablecaption{The medium value with the 16\% and 84\% percentile error of the posterior distribution for the parameters in models M1--M3. \label{tab:had_lep_param}}
\tablecolumns{7}
% \tablenum{2}
\tablewidth{0pt}
\tablehead{
  %%%% header 1st line
  \colhead{Model} &
  \colhead{$\alpha_1$} &
  \colhead{$\alpha_2$} & 
  \colhead{$p_\mathrm{b}$} & 
  \colhead{$p_\mathrm{c}$} & 
  \colhead{$\alpha_{\rm e/p}$} &
  \colhead{$p_\mathrm{c,e/p}$} & 
  \colhead{$W_\mathrm{2,e/p}$} &
%  \colhead{$B$} &
  \colhead{BIC} \\
  %%%% header 2nd line
  & & & ($\mathrm{GeV\ c^{-1}}$) & ($\mathrm{TeV\ c^{-1}}$) &  & ($\mathrm{TeV\ c^{-1}}$) & ($10^{48}$ erg) 
  %& ($\uG$)
}

\startdata
%  M0 & $2.03^{+0.01}_{-0.02}$ & -- & -- & $9.0^{+1.1}_{-1.0}$ & -- & -- & -- & -- & 131.6 \\
 M1 & $1.0^a$ & $2.29^{+0.04}_{-0.04}$ & $14.0^{+2.4}_{-2.1}$ &  $28.3^{+8.7}_{-6.7}$  & -- & -- & -- & 59.8\\
 M2 & $1.0^a$ & $3.05^{+0.60}_{-0.26}$ & $34.2^{+11.2}_{-6.5}$ & 3000$^a$ & $<1.57$ & $3.0^{+1.6}_{-0.7}$ & $33.7^{+9.2}_{-7.0}$  & 41.9 \\
 M3 & $1.0^a$ & $3.03^{+0.67}_{-0.23}$ & $32.8^{+12.3}_{-6.6}$ & 3000$^a$ & $1.20^{+0.70}_{-0.57}$ & $1.3^{+0.6}_{-0.4}$ & $0.3^{+0.4}_{-0.1}$  & 43.4 \\
%  M5 & $1.0^a$ & $2.82^{+0.25}_{-0.13}$ & $33.3^{+8.8}_{-5.2}$ & 3000$^a$ & $2.49^{+0.01}_{-0.01}$ & $9.4^{+0.4}_{-0.6}$ & $2.7^{+0.4}_{-0.5}$ & $143.9^{+16.2}_{-10.7}$ & -- \\
\enddata
\tablenotetext{a}{Fixed in the fitting process.}
% \tablecomments{}
\end{deluxetable}

\end{document}

%% file: authors.tex
 
\author{Zhen Cao}
\affiliation{Key Laboratory of Particle Astrophysics \& Experimental Physics Division \& Computing Center, Institute of High Energy Physics, Chinese Academy of Sciences, 100049 Beijing, China}
\affiliation{University of Chinese Academy of Sciences, 100049 Beijing, China}
\affiliation{TIANFU Cosmic Ray Research Center, Chengdu, Sichuan,  China}
 
\author{F. Aharonian}
\affiliation{TIANFU Cosmic Ray Research Center, Chengdu, Sichuan,  China}
\affiliation{University of Science and Technology of China, 230026 Hefei, Anhui, China}
\affiliation{Yerevan State University, 1 Alek Manukyan Street, Yerevan 0025, Armenia}
\affiliation{Max-Planck-Institut for Nuclear Physics, P.O. Box 103980, 69029  Heidelberg, Germany}
 
\author{Y.X. Bai}
\affiliation{Key Laboratory of Particle Astrophysics \& Experimental Physics Division \& Computing Center, Institute of High Energy Physics, Chinese Academy of Sciences, 100049 Beijing, China}
\affiliation{TIANFU Cosmic Ray Research Center, Chengdu, Sichuan,  China}
 
\author{Y.W. Bao}
\affiliation{Tsung-Dao Lee Institute \& School of Physics and Astronomy, Shanghai Jiao Tong University, 200240 Shanghai, China}
 
\author{D. Bastieri}
\affiliation{Center for Astrophysics, Guangzhou University, 510006 Guangzhou, Guangdong, China}
 
\author{X.J. Bi}
\affiliation{Key Laboratory of Particle Astrophysics \& Experimental Physics Division \& Computing Center, Institute of High Energy Physics, Chinese Academy of Sciences, 100049 Beijing, China}
\affiliation{University of Chinese Academy of Sciences, 100049 Beijing, China}
\affiliation{TIANFU Cosmic Ray Research Center, Chengdu, Sichuan,  China}
 
\author{Y.J. Bi}
\affiliation{Key Laboratory of Particle Astrophysics \& Experimental Physics Division \& Computing Center, Institute of High Energy Physics, Chinese Academy of Sciences, 100049 Beijing, China}
\affiliation{TIANFU Cosmic Ray Research Center, Chengdu, Sichuan,  China}
 
\author{W. Bian}
\affiliation{Tsung-Dao Lee Institute \& School of Physics and Astronomy, Shanghai Jiao Tong University, 200240 Shanghai, China}
 
\author{A.V. Bukevich}
\affiliation{Institute for Nuclear Research of Russian Academy of Sciences, 117312 Moscow, Russia}
 
\author{C.M. Cai}
\affiliation{School of Physical Science and Technology \&  School of Information Science and Technology, Southwest Jiaotong University, 610031 Chengdu, Sichuan, China}
 
% ...
%\collaboration{The LHAASO Collaboration}